# Silicon Hard-Stop Spacers for 3D Integration of Superconducting Qubits

Bethany M. Niedzielski[1], David K. Kim[1], Mollie E. Schwartz[1], Danna Rosenberg[1], Greg Calusine[1], Rabi Das[1], Alexander J. Melville[1], Jason Plant[1], Livia Racz[1], Jonilyn L. Yoder[1], Donna Ruth-Yost[1], William D. Oliver[1, 2]
[1]MIT Lincoln Laboratory, 244 Wood Street, Lexington, MA, 02420, email: Bethany.Huffman@ll.mit.edu, David.Kim@ll.mit.edu
[2]Department of Electrical Engineering & Computer Science, Massachusetts Institute of Technology, Cambridge, MA, USA

*Abstract—* As designs for superconducting qubits become more complex, 3D integration of two or more vertically bonded chips will become necessary to enable increased density and connectivity. Precise control of the spacing between these chips is required for accurate prediction of circuit performance. In this paper, we demonstrate an improvement in the planarity of bonded superconducting qubit chips while retaining device performance by utilizing hard-stop silicon spacer posts. These silicon spacers are defined by etching several microns into a silicon substrate and are compatible with 3D-integrated qubit fabrication. This includes fabrication of Josephson junctions, superconducting air-bridge crossovers, underbump metallization and indium bumps. To qualify the integrated process, we demonstrate high-quality factor resonators on the etched surface and measure qubit coherence ($T_1$, $T_{2,echo}$ > 40 µs) in the presence of silicon posts as near as 350µm to the qubit.

## I. Introduction

As quantum computing with superconducting qubits advances past the few-qubit stage, it has become necessary to go beyond planar geometries to enable larger-scale circuits. Heterogeneous 3D integration has been shown to be a viable pathway to meet the increased interconnect needs of these systems, while preserving qubit performance [1]. Fig. 1 shows a top-down and cross-sectional view of MIT Lincoln Laboratory's three-tier stack architecture, which uses In bump-bonding to combine a qubit tier with high-coherence qubits, an interposer tier with two metal layers connected with superconducting through-silicon vias [2], and a multichip module tier with multiple superconducting routing layers [3]. This multi-tier approach facilitates the connectivity and addressability needs of a large qubit system, while shielding the qubits from unnecessary processing steps by independently fabricating each tier.

In a multi-chip architecture, the tilt and spacing between the qubit and interposer tiers must be tightly controlled. Any tilt causes variations in the electromagnetic environment across the chip, which will impact the inductive and capacitance coupling between chips and alter the qubit and resonator frequencies through capacitive loading. Standard bump-bonding techniques used with superconducting qubits produce tilts of 500 µrad across square 4 - 6 mm chips [4]. Building complex circuits with many qubits and dense resonator frequency multiplexing will require tighter control of chip spacing to enable precise targeting and avoid frequency collisions. To achieve this goal, we implement hard-stop Si spacer posts etched directly into the qubit chip. The etch uniformity and post heights dictate the spacing and tilt between the qubit and interposer tiers following the bump bonding process.

## II. Fabrication and Characterization

Starting with a factory-polished, 50 mm Si wafer, the hard-stop spacer posts are defined by masking locations where these posts are desired using 10 µm of NR9-8000 negative photoresist with a hexamethyldisilazane (HMDS) adhesion promoter and performing a subtractive etch. The spacer posts are 100 x 100 µm$^2$ in size and positioned 700 µm from each corner of the 5 x 5 mm$^2$ chips tiled across the wafer. The bulk of the Si wafer is etched down several microns using an inductively coupled plasma (ICP) etch with 20 sccm $Cl_2$, 3 sccm $BCl_3$, 1.5 Pa chamber pressure, 50 W bias power, and 300 W ICP power at 50° C. Etches of 2 - 4 µm were demonstrated with an etch rate of 170 nm/min. The uniformity of this etch determines the planarity of subsequently bonded devices.

An atomic force microscopy image of a blanket-etched wafer is shown in Fig. 2, where the targeted etch depth was 2 µm. Compared to the factory polished RMS roughness of 0.15 nm, an RMS roughness of 0.3 nm was measured on the etched Si surface. A sample set of 15 Al resonators fabricated on top of the etched surface was measured with an average internal quality factor ($Q_i$) of 800,000 at single-photon power levels [5]. This is comparable to our reference sample that was fabricated on top of a factory polished Si surface with a $Q_i$ of 825,000, indicating that the etched surface is compatible with high-quality superconducting films.

To verify that the etched post height uniformity was sufficient for reducing bump-bonding tilt, we used confocal microscopy to measure post height vs. radius for over 100 spacer posts across a 50 mm wafer (Fig. 3a). A 3 µm etch depth was targeted and an average post height of 3.13 µm was achieved, which is indicated in the plot with a horizontal black line. Outside of a 5 mm edge exclusion due to tool specifications, the measured height is within 5% of the target height. In addition, the data is tightly clustered about the average, giving an etch uniformity of σ = 2.4%. Based on the height variation of the four corner posts, the tilt of every 5 x 5 mm$^2$ chip was calculated across the wafer (Fig. 3b). This was repeated for hundreds of chips over dozens of wafers, yielding an average calculated chip tilt of 11 ± 7 (1σ) μrad. Even the maximum calculated tilt from all measured chips of 86 μrad was well below the reported tilt values of 500 μrad without spacers.

### III. PROCESS INTEGRATION

Capacitively shunted flux qubit (CSFQ) devices were fabricated around the Si spacer posts and on top of the etched Si surface in order to validate their compatibility with qubit fabrication [6]. High-$Q_i$ Al was patterned into control and readout circuitry and capacitive shunts, followed by the addition of Josephson junctions [6], air-bridge crossovers [2,7], under-bump metallization (UBM), and In bumps [1]. For each process step, the post tops needed to be either protected with resist or cleaned of any deposited material in order to retain their planarity. We implemented both techniques by increasing resist thicknesses and adding a step to etch residual material from the post tops using Transene Al etchant Type A.

Scanning electron microscopy (SEM) images of a completed wafer shows a spacer post surrounded by UBM pads patterned with In bumps in Fig. 4a, with a resonator line and air-bridge crossovers in Fig. 4b, and 350 μm from a qubit loop in Fig. 4c. The images demonstrate that these elements can be successfully fabricated in close proximity to the spacer post without degradation of their physical dimensions. In addition, the tops of the spacer posts remained clean after this processing. The room temperature resistance of single feature and cross bridge Kelvin resistance (CBKR) test structures also showed no change from the etched Si surface or the presence of posts.

A qubit chip with Si spacer posts was then bump-bonded to an interposer chip to verify the improvement in tilt using SEM. The edge of the bonded sample was ground back to reveal the posts, as shown in Fig. 5. The left side of the image and its inset show that the post on the qubit chip makes complete contact with the Al layer on the interposer chip, thereby determining the spacing between the two chips. The three In bumps on the right side of the image and its inset were compressed down to the height of the posts during the bonding process and also make complete contact with the interposer chip. Daisy chain measurements demonstrated a high-yield, low-resistance galvanic connection between the chips at temperatures well below 1K. This chain consisted of 2704 bump connections snaking between the qubit and interposer chips and supported a critical current greater than 10mA.

### IV. QUBIT MEASUREMENTS

The final Si spacer validation needed to enable use in a quantum processor is to show that qubits integrated with hard-stop posts maintain their high coherence properties. To this end, qubits were co-fabricated with Si spacer posts to determine how close a qubit could be placed to a post. Room temperature resistance measurements were first used to address this question. A test-structure block was measured with 85 nominally identical 200 x 200 $nm^2$ Josephson junctions centered around a spacer post. Fig 6. shows the normalized resistance of the junctions against their distance from the spacer post for two samples with posts and a control sample without posts. The resistance values have no dependence on the distance from a spacer post and exhibit 2-3% cross-wafer resistance variation, which is consistent with our standard process [6]. SEM inspection of the junctions show that their size and shape were not impacted when patterned as close as 350 μm from the post.

Cryogenic measurements of Al CSFQ devices were performed on chips with four corner spacer posts and a fifth post positioned from 350 to 1500 μm from the center of the qubit loop. Measurements of qubit coherence times showed no discernable difference between qubits with no spacer posts, four spacer posts, or with a fifth post 350 μm from the qubit loop. A representative coherence time measurement is plotted in Fig. 7 with $T_1$ = 45 μs and $T_{2, Echo}$ = 55 μs.

### V. CONCLUSION

A process for hard-stop Si spacer posts was developed to control the tilt and spacing between 3D integrated tiers for superconducting qubit applications. These posts were reliably etched into over 100 chips with a uniformity corresponding to 11 ± 7 (1σ) μrad of tilt, well below that of bump bonding without spacer posts in superconducting qubit systems. The spacers were successfully integrated into the full qubit fabrication flow in close proximity to qubits, superconducting air-bridge crossovers, UBM, and In bumps, with demonstrated high qubit performance. The spacers enable dense 3D integrated qubit designs with tightly controlled spacing and tilt between bump-bonded chips, while preserving high qubit coherence, as we build toward larger-scale systems.

### ACKNOWLEDGMENTS

We gratefully acknowledge M. Augeri, P. Baldo, G. Fitch, M. Hellstrom, J. Liddell, K. Magoon, P. Murphy, B. Osadchy, C. Thoummaraj and D. Volfson at MIT Lincoln Laboratory for technical assistance. This research was funded by the ODNI and IARPA. The views and conclusions contained herein are those of the authors and should not be interpreted as necessarily representing the official policies or endorsements, either expressed or implied, of ODNI, IARPA, or the US Government.

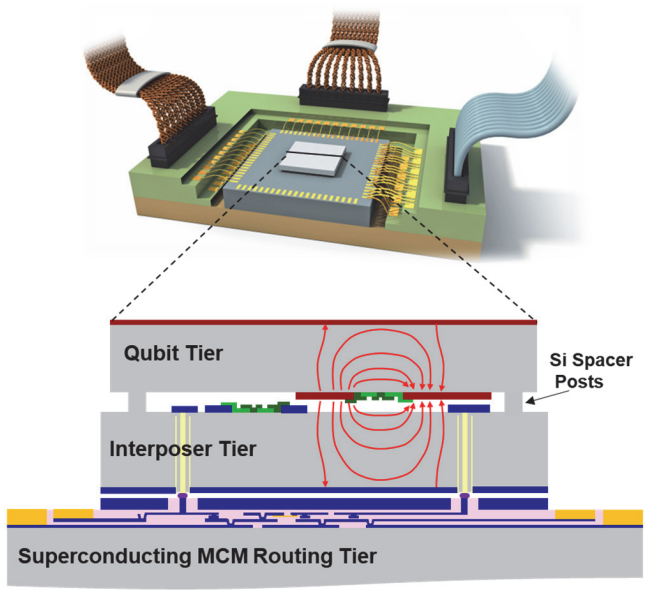

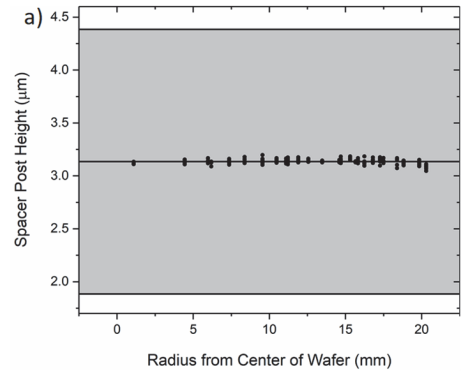

Fig 1: *Three tier stack architecture for superconducting qubit systems.* This design features three separately fabricated device tiers to enhance qubit connectivity while retaining high coherence. The individual tiers are mechanically and electrically connected through In bump bonding with Si spacer posts to control the tilt and spacing between the qubit and interposer tiers. From Ref. [1].

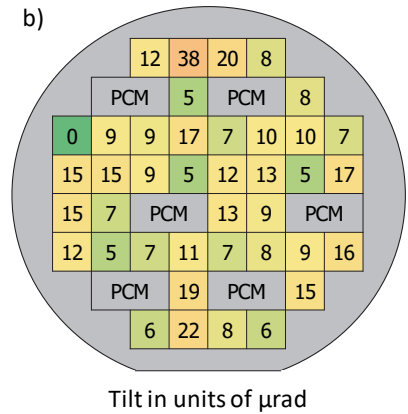

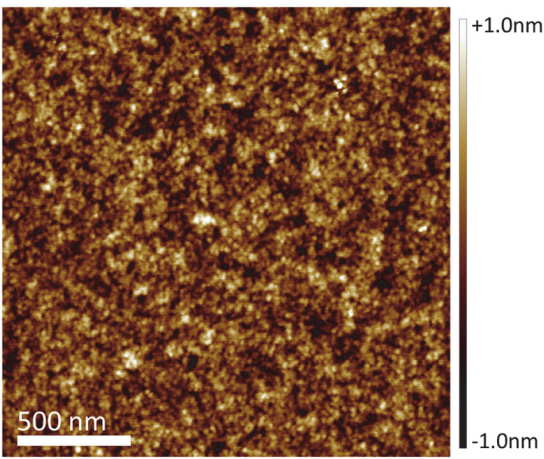

Fig 2: *AFM image of the etched Si surface.* The RMS roughness of the etched Si surface was less than 0.3 nm. Prior to etching, the factory polished Si surface had an RMS roughness of 0.15 nm.

Fig 3: *Sample set of etched Si spacer post heights from representative wafer.* a) Height variations of over 100 spacer posts across a 50 mm wafer with a 5 mm edge exclusion. A 3 μm post height was targeted and the horizontal black line shows the average height of 3.13 μm. The gray shaded region corresponds to the height variation needed for 500 μrad of tilt between bonded chips, which has been achieved by In bump bonding for superconducting qubits without spacer posts. b) By grouping the four spacer posts on each chip, the local height variation of those posts can be used to calculate the resulting tilt for each chip. For this representative wafer, the average calculated chip tilt was $11 \pm 6$ (1σ) μrad. The 12 chips without tilt values are process control monitor (PCM) chips that were not etched with corner posts.

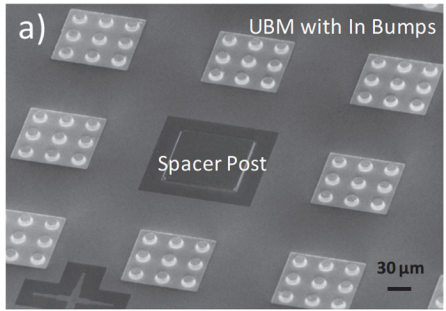

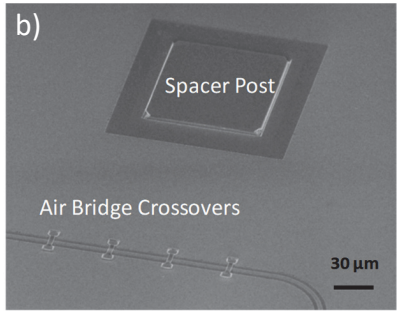

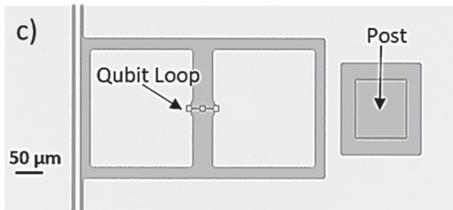

Fig 4: *SEM images of process integration with Si spacer posts.* The tops of the Si spacer posts are clean and feature shapes and sizes are preserved after metal layer, Josephson junction, air-bridge crossover, underbump metal (UBM), and In bump patterning in close proximity to spacer posts.

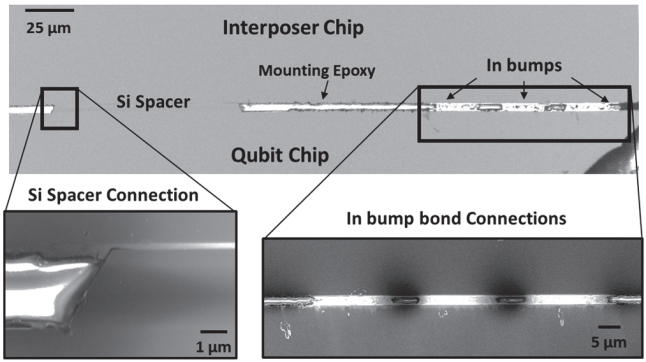

Fig 5: *Cross-Sectional SEM of a bump-bonded qubit and interposer chip with Si spacer posts.* The insets show that the posts and In bumps make good contact between the qubit to interposer chips, so the heights of the posts will set the spacing and tilt of the bump-bonded device. These chips were filled with mounting epoxy only to preserve the structure during edge grind to reveal this surface.

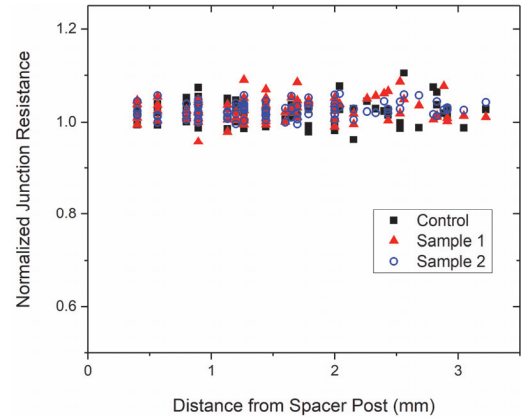

Fig 6: *Room temperature resistance of Josephson junction test structures fabricated near Si spacer posts.* This plot shows the four-point room temperature resistance of 85 nominally identical Josephson junctions with respect to their distance from a Si spacer post. Data from a control sample with no spacer post and two samples with a spacer post in the test structure block show no discernable resistance difference with post proximity.

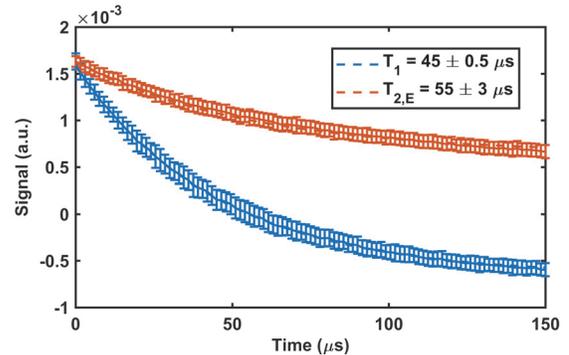

Fig 7: *Lifetime and coherence measurements of qubits with Si spacer posts.* Measurements of qubit lifetime, $T_1$, and phase coherence time, $T_{2,E}$ (Hahn echo), were comparable for samples with no posts, four corner posts, and a fifth post 350 - 1500 μm from the qubit. The data shown is from a representative sample with four corner posts, which is our standard layout.